\documentstyle[epsf]{l-aa}
\def\apj{ApJ}
\def\apjs{ApJS}
\def\aap{A\&A}
\def\aas{A\&AS}
\def\aj{AJ}
\def\mn{MNRAS}
\def\pasp{PASP}

\begin{document}
\thesaurus{06(08.05.3, 10.06.1, 10.07.2, 10.08.1)}
\title{Chronology of the halo globular cluster system formation}
\author{M.~Salaris\inst{1,2} \and A.~Weiss\inst{1}}
\institute{Max-Planck-Institut f\"ur Astrophysik,
Karl-Schwarzschild-Str.~1, 85740 Garching, Germany
\and
Institut d'Estudis Espacials de Catalunya, 08034 Barcelona, Spain
}
\offprints{M.~Salaris (e-mail address: maurizio@mpa-garching.mpg.de)}
\maketitle

\begin{abstract}
Using up-to-date stellar models and isochrones we determine the
age of 25 galactic halo clusters. The clusters are distributed into
four groups according to metallicity. 
We measure the absolute age of a reference cluster in each group, and
then find the relative ages of the other clusters relative to this
one. 
This combination yields the most reliable
results. We find that the oldest cluster group on average is
$11.8\pm0.9$ Gyr or $12.3\pm0.3$ Gyr old, depending on whether we
include Arp2 and Rup106. The average age of all clusters is about 10.5
Gyr. Questions concerning a common age for all clusters and a
relation between metallicity and age are addressed. The  groups
of lower metallicity appear to be coeval, but our results
indicate that globally the sample has an age spread without a simple
age-metallicity relation. 
\end{abstract}

\section{Introduction}

The age of galactic Globular Clusters (GCs) provides fundamental
information about the age of the
universe and the formation history of the Galaxy.
Recent improvements in the input physics needed for computing 
stellar evolutionary models
have revived theoretical work on low-mass stars 
and GC age determinations (Chaboyer \& Kim 1995, 
Mazzitelli et al.\ 1995, D'Antona et
al. 1997, Salaris et al.\ 1997). Salaris et
al. (1997 - Paper~I) 
have shown that the age of the supposedly oldest clusters -- the most
metal-poor ones -- is around 12 Gyr. This age reduction with respect to
earlier work (e.g.\ Chaboyer et al.\ 1992, 
Salaris et al.\ 1993)
has been identified to be mainly due to the use of
an improved equation of state that
includes non-ideal gas effects (Rogers et al.\ 1996). 

After this initial work, which was motivated by a possible ``conflict
over the age of the universe'', the next step is to address
the question of Galaxy formation. This means that one has to investigate many
clusters and determine their ages, looking for correlations of age with
cluster metallicity or galactocentric distance.

The position of the turn-off (TO) is the feature
in the colour-magnitude-diagrams (CMD) of stellar clusters that is most
sensitive to the age of the stellar population. The higher the cluster age,
the less luminous and redder is the TO. Two differential quantities
are suited as age indicators that are 
independent of reddening and distance: the brightness difference in $V$
magnitude between 
the TO and the horizontal branch (HB)
at the RR Lyrae stars region, called the $\Delta(V)$ or vertical method
(see, e.g., Sandage \& Cacciari 1990; Stetson et al.\ 1996
for a review) 
and the $(B-V)$ colour difference between the TO and the base of the
Red Giant Branch (RGB), called the $\Delta(B-V)$ or horizontal method
(see, e.g., Chieffi \& Straniero 1989, VandenBerg et al.\ 1990).
In both
cases the TO position is differentially determined with respect to a
CMD branch (the HB or the RGB) whose location is virtually 
independent of age in the case of old stellar populations.

Direct absolute age determinations based on the vertical method lead to
a large spread in ages, which seems to be correlated with
metallicity (Chaboyer et al.\ 1996). However, the quality and
diverseness of the data do not favour this method. The horizontal
method works best if 
all problems with theoretical effective temperatures (convection
theory, atmospheres) and the conversion of theoretical quantities to
observed colour and brightness are avoided. It therefore is very well
suited for obtaining relative ages of clusters of similar metallicity,
but absolute ages are very difficult to obtain.

A combination of both methods seems therefore to be promising for an
accurate determination of absolute and relative ages.
Clusters are inspected in groups of similar metallicity, and one or
more suitable ``template'' clusters 
(with homogeneous and good photometry for both TO and HB region)
are chosen to determine an
absolute age directly with the vertical method. Then, the horizontal
method is used for a differential comparison with other  
clusters of the same group.  This is the approach we have chosen (see
also Paper~I) and it is similar to the one by
Richer et al.\ (1996).  
Our work differs from their analysis in that we use the
vertical method for determining the absolute ages by virtue of theoretical
isochrones and zero-age horizontal-branch models, while their absolute
ages were obtained 
by fitting Bergbusch \& VandenBerg (1992) isochrones, 
without explicitly using theoretical HB
models; moreover we use new and improved stellar models
(see Paper~I), taking into account the latest developments regarding
stellar opacities and equation of state.

The questions we want to address are: (i) what is the absolute age of
one or more template cluster in each metallicity group; (ii) how do
$\Delta(B-V)$ differences between clusters within a group
translate into age differences? This information is necessary for the
more global problem of whether there is an age spread between
galactic clusters, and if so, whether it is correlated with
metallicity.  Assessing clearly the existence of an age spread among
GCs and of an age-metallicity relation is fundamental for
understanding the formation of our Galaxy.  Very recently
Chaboyer et al.\  (1996) found strong evidence in favour of a spread in the ages
of galactic GCs and an age-metallicity relationship.
On the contrary, Stetson et al.\ (1996) concluded that there is no evidence for a
significant spread in ages among clusters of similar metallicity and
that the case concerning age differences between metallicity groups
remains unsettled, while Richer et al.\ (1996) found that the most
metal-poor clusters may be slightly older than clusters of higher
metallicities. The results of the latter group are, however, neither
inconsistent with a 
picture in which all clusters of all metallicities formed
simultaneously. Between the most metal-rich clusters there appears to
be a considerable age spread of $\sim 2 $ Gyr (VandenBerg et al.\
1990), and in addition a number of exceptional clusters were found in
all investigations. 

In the present paper we restrict ourselves to halo clusters - that
means GCs with kinematic properties typical of the halo component of
the Galaxy - 
and try to answer both the questions concerning the distribution of
ages within individual metallicity groups and between groups.  
In Sect.~2 we will review our
method used for determining absolute and relative cluster
ages. Sect.~3 contains our results for a large group of halo clusters,
which are compared  with the results by Richer et al.\ (1996). In
Sect.~4 the implications for the cluster age distribution
are discussed, and a summary of our results follows in the final section.

\section{Method}

\subsection{Stellar Models}

As in Paper~I we rely on theoretical models for all evolutionary
phases; in particular we have computed stellar models from
the main-sequence (MS) up to the zero-age horizontal branch (ZAHB).
The input physics 
employed in the models is the same as in Paper~I: for the
opacities we used a combination of the latest OPAL opacities (Rogers
\& Iglesias 1992; Iglesias \& Rogers 1996) and tables from
D.~Alexander (Alexander \& Ferguson 1994; Alexander, private
communication). The metal mixtures included identical $\alpha$-element
enhancement for all opacity tables. 
The equation of state consisted of the OPAL EOS (Rogers,
Swenson \& Iglesias 1996) with extensions for the lowest temperatures
and degenerate helium cores taken respectively from Chieffi \&
Straniero (1989) and Straniero (1988).

Diffusion of helium and heavy elements is not included in the calculations. 
The effect of diffusion on the ages obtained
from our models is discussed in  Sect.~2.4, which is dedicated to the methods
for determing absolute and relative ages. 

We computed stellar models for the following compositions: $(Z,Y)$ =
(0.0002, 0.230), (0.0004, 0.230), (0.0006, 0.232)
(0.0008, 0.232), (0.001, 0.233), (0.0015, 0.235)
(0.002, 0.236), (0.004, 0.242). For $\triangle Y / \triangle Z$ we
have taken a mean value of 3 (as in Bergbusch \& VandenBerg 1992;
Mazzitelli et al.\ 1995).  The first two mixtures are those already used
in Paper~I. $\alpha$-elements are always enhanced (e.g.\ $[{\rm O/Fe}]
= +0.5$); the total metal-abundance $[{\rm M/H}]$ is about
0.2-0.3 dex higher than $[{\rm Fe/H}]$.

Stellar models with masses between 0.7 and $1.0 M_\odot$ were evolved
from the zero-age main sequence 
up to the RGB.  The mixing length has been calibrated as
explained in Paper~I.  Isochrones for different ages were computed
from these evolutionary models.  ZAHB models with varying envelope
masses were calculated as described in Paper~I.

Similarily, the conversion from theoretical effective temperature and
luminosity to observable colour $(B-V)$ and visual magnitude $V$ was
done using the transformations of Buser \& Kurucz (1978, 1992).

\subsection{Globular Clusters groups}

In this paper (as in Paper~I) we have used both the vertical and the
horizontal method for determining the distribution of the halo GCs
ages.
The strengths and weaknesses of these techniques have been
discussed extensively in many recent papers (VandenBerg et
al.\ 1990, Salaris et al.\ 1993, Chaboyer et al.\
1996). To summarize, the $\Delta(V)$ appears particularly
suitable for the determination of the absolute clusters ages, since
it is independent of the treatment of convection and is only weakly
sensitive 
to metallicity. However, it can be safely applied only in the case of
clusters with homogeneous 
photometry for both TO and HB, and with a well populated
horizontal part of the HB. Clusters with only a blue, vertical HB
are in principle excluded,  because it is impossible to have
observational estimates of the absolute HB luminosity in the RR
Lyrae region or to constrain the age from the fit of theoretical ZAHB
models.  

The $\Delta(B-V)$ can in principle be
applied to each kind of GC with sufficiently accurate photometry,
since each GC shows a main-sequence TO and an RGB. It is
only weakly sensitive to metallicity, but  
absolute ages depend on the mixing length calibration and
on the transformations from  effective temperatures
to colours.  When all the problems related to the $T_{\rm eff}$
determination and the conversion to observed colours are minimized, as
in the case of clusters with similar metallicities, the horizontal
method turns out to be suitable for accurately determining the
relative ages of clusters (VandenBerg et al.\ 1990, Stetson et al.\ 1996).

Since our goal is to study the distribution of ages of a well
populated sample of halo GCs, one has to deal with clusters with very
different HB types and with photometries not always extended up to the
HB, or showing only a scarcely populated or blue HB.
Therefore, we have to apply a combination of both methods for getting
the ages of the cluster sample.

We have collected published CCD data for 25 halo clusters
(see Table~1), which span a wide range of metallicities,
galactocentric distances and HB types, divided into four groups according
to their metallicity.
The first group spans the range
$-2.1\leq{\rm [M/H]}<-1.6$ (metal poor clusters), the second
$-1.6\leq{\rm [M/H]}<-1.3$ (intermediate metal poor clusters), the third
$-1.3\leq{\rm [M/H]}<-0.9$ (intermediate metal rich clusters) and the
fourth $-0.9\leq{\rm [M/H]}<-0.6$ (metal rich clusters).  The adopted
$\rm [Fe/H]$ values come from Zinn (1985), and the global metallicity has
been obtained considering an average $[\alpha{\rm/Fe}]=0.3$ and
${\rm [M/H]}={\rm[Fe/H]}+0.2$ according to the discussion in Paper~I. The
metallicity difference among the clusters in each group is about a
factor of 2.  In the case of Rup106, Arp2 and Ter7, for which there
are no data in Zinn (1985), we have considered the $\rm[Fe/H]$ estimates
by Buonanno et al (1993, 1995a, 1995b) to which we have added the
contribution of the $\alpha$-elements
\footnote{Carretta \& Gratton (1997) recently have
recalibrated the Zinn \& West (1984) $\rm [Fe/H]$ scale, which corresponds to
the Zinn (1985) metallicity scale for almost all GCs in our sample. 
The new calibration is based on a homogeneous set of cluster $\rm
[Fe/H]$ determinations from high resolution spectroscopic data; it provides 
$\rm [Fe/H]$ values for the clusters in our sample 
that are on average 0.20 dex higher than the ones we used.
This simply corresponds to an almost constant shift 
of the $\rm [M/H]$ ranges adopted for our GCs groups, 
but the membership of each single group does not change 
and similarly the absolute and relative ages we derive are only very
marginally affected by the choice of this new scale (see 
Sect.~3, which is about age determination).},
assuming that such an enrichment exists for these clusters as well.

We have considered only clusters with photometries that show at least
MS, TO and RGB, and that permit a clear determination
of the TO position (within an error $\leq \pm$0.15 mag). 
In each group one cluster (or two clusters, if possible) is selected as 
the ``reference'' one, and its absolute age is determined directly
by means of the vertical method; the photometry of the reference cluster
has to show not only a clearly defined TO position, but also 
the RGB and a well-populated HB of such morphology that the ZAHB level
can be safely determined.
Thus, CMD morphology is more important than the
overall quality of the photometry for a cluster to be suitable as a
reference cluster.

Within each group the
relative ages with respect to the reference cluster are determined by
means of the horizontal method. Where there are two reference clusters the
age difference from the vertical method can be cross-checked with that
derived by means of the horizontal one.

\subsection{Absolute age determination}

The advantage of using the vertical method for determining the
absolute age of a cluster is that it is largely independent
of all uncertainties 
connected with the calculation of effective temperatures and their
conversion to colours. It does not depend on the reddening and on the
assumed distance modulus (the same holds, of course, also for the
horizontal method), although the models yield a distance scale (by
means of the comparison between observed and theoretical ZAHB level),
which can be compared to independently determined values.

To obtain the
cluster age, the procedure is the following: from the observed
brightness of HB stars the apparent ZAHB brightness is derived (see
below); with the TO-brightness as given by the observers, $\Delta V$
is determined uniquely and is compared to theoretical predictions
of $\Delta V$ as a function of age for isochrones and ZAHB models of the
appropriate metallicity. These steps are sufficient to find the
cluster age.  

However, one can go beyond this to check the reliability of the
results. First, the difference between observed apparent and
theoretical absolute ZAHB brightness gives the distance modulus
following from our models, which can be compared to independent
determinations from other distance
indicators; thus, the distance modulus provides an independent
way to assess the reliability of the derived ages.

Second, the isochrone can be compared to the CMD. Since
age and $(m-M)_V$ already have been fixed, isochrones (and ZAHB models)
remain to be shifted in
colour to match the observed MS. This shift corresponds to the
cluster reddening $E(B-V)$.
The overall fit to the
observed CMD may serve as an additional qualitative indicator. There is,
however, the following point to be taken into account: 
since our age determination method does not rest on detailed isochrone
fitting, we have taken 
the metallicity directly from the literature, without trying
to improve the fit by exploring the error range in [M/H]. 
Changing the metallicity within the allowed range can improve the
isochrone fit (see Sect.~3.2) without affecting appreciably the
determined age.

An important step in the vertical method is therefore to fix the zero-age level
of the observed HB. In Paper~I we selected two clusters (M68 and M15) with
well-populated HBs and a sufficiently large number of RR~Lyrae
variables. 
With the mean brightness ($\langle V_{\rm RR}\rangle$) of
the variables fixed, we derived the zero-age level using the relation
by Carney et al.\ (1992):

\begin{equation}
V_{\rm ZAHB}=\langle V_{\rm RR}\rangle+0.05[{\rm Fe/H}]+0.20
\label{vzahb}
\end{equation}

\begin{figure}
\begin{center}
  \mbox{\epsfxsize=0.9\hsize\epsffile{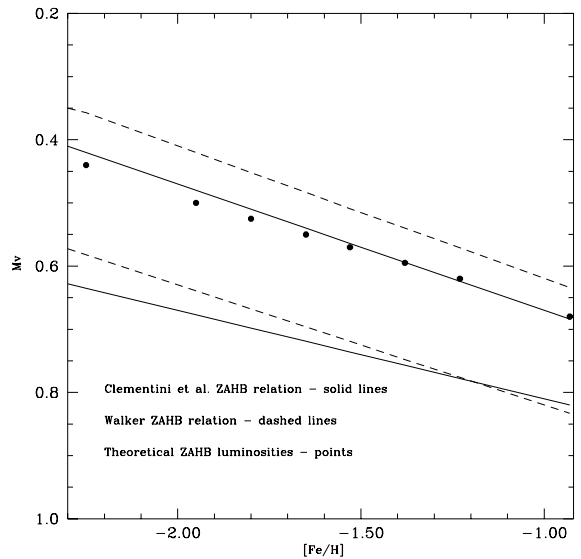}}
\end{center}
\caption{Theoretical ZAHB luminosities as a function of $\rm[Fe/H]$.
Solid and dashed lines represent the upper and lower envelopes of the
empirical determinations by Clementini et al.\ (1995) and Walker
(1992a)}
\label{theorhb}
\end{figure}

We have also demonstrated in Paper~I 
that (assuming a constant helium content of $Y=0.23$) our theoretical
relation between ZAHB brightness (taken at $\log T_{\rm eff}=3.85$) and
$[{\rm Fe/H}]$ together with Eq.~\ref{vzahb} agrees nicely with
that found empirically by Clementini et al.\ (1995) and also supports
that of Walker (1992a). The reader is referred to Paper~I
for a deeper discussion about the problem of the 'true'
observational relation between RR~Lyrae luminosities and metallicity.
In the present paper we consider an initial helium
abundance varying with $Z$, but the difference with the ZAHB
luminosities obtained in Paper~I is always less than 0.04
mag (reached at the highest metallicities considered).  
In Fig.~\ref{theorhb} we show the final
relation between ZAHB luminosity and $[{\rm Fe/H}]$ from our
stellar models; in the same figure the
empirical relations  by Clementini et al.\  (1995 - when considering
only ZAHB objects) and Walker (1992a) are also displayed (upper and
lower limits according to the errors given by the authors). The
agreement between the ZAHB theoretical models and both empirical
determinations is evident.

Very recently, first results on the absolute luminosity of HB
(Feast \& Catchpole 1997; de Boer et al.\ 1997) and
subdwarfs stars (Reid 1997; Gratton et al.\ 1997)
based on HIPPARCOS data have appeared. The implications of the latter
papers for the distances to M68 and M5 we will discuss in the
respective sections. 
Feast \& Catchpole (1997)
used the trigonometric parallaxes of galactic Cepheid variables for
recalibrating the zero-point of the period-luminosity relation. After
applying a correction for metallicity effects, they derive a distance
modulus of $18.70\pm0.10$ for the LMC, almost 0.20 mag higher than the 
value commonly used, based on a previous Cepheids calibration and on
other distance indicators (see, e.g. Walker 1992a).
This distance modulus, coupled with the RR Lyrae observations by
Walker (1992a) for LMC clusters, provides a mean absolute magnitude
$\langle V_{\rm RR}\rangle =0.25\pm0.10$ at
[Fe/H]=-1.9 for RR Lyrae stars; this value corresponds, by applying
Eq.~\ref{vzahb}, to $V_{\rm ZAHB}=0.36\pm0.10$. From our theoretical
models we get an 
absolute  ZAHB luminosity $V_{\rm ZAHB}=0.51$ at the instability strip
for the same metallicity. This means that our theoretical value would
be higher than the upper limit of the observationally 
allowed range (0.46) by 0.05 mag.  

The opposite conclusion can be reached when considering the results by
de~Boer et al.\ (1997), again based on HIPPARCOS parallaxes; 
they present absolute $V$ magnitudes for a group of HB field halo stars, 
bluer than the instability strip. We consider for example
the reddest star in their sample (HD161817), which is also the object
with the smallest error on $M_{V}$ (the errors for the other stars are
much bigger, up to 1.25 mag). The observations give $M_{V}=0.72\pm0.25$ and
$(B-V)_{0}$=0.14. The metallicity of the objects is not
given in the paper. By using our theoretical ZAHB models with ${\rm
[Fe/H]}=-1.03$ 
(as done by the authors when comparing their data with the ZAHB models
by Dorman 1992) we obtain $M_{V}=0.67$ at $(B-V)_{0}$=0.14. If we
consider the models with ${\rm[Fe/H]}=-1.6$, which corresponds
approximately to the average metallicity of field halo stars, we
obtain $M_{V}=0.58$. Both values are in
agreement with the observationally allowed range.
If, however, we take into account the
fact that HD161817 is likely to be evolved from the ZAHB towards higher
luminosities (as also noted by de~Boer et al.\ 1997), then the theoretical 
models are more luminous than observationally allowed.
Therefore, the two papers discussed lead to discrepant
results. Compared to Feast \& Catchpole (1997) our predicted distance
moduli appear to be too low; compared to de~Boer et al.\ (1997) they
are too large. The first results based on HIPPARCOS data do not 
disprove the reliability of our theoretical ZAHB luminosities.

To determine the observational ZAHB brightness, the application of
Eq.~1 requires good HB photometry and the presence of a sufficient
number of RR~Lyr stars. It was not possible for all metallicity groups
to find a cluster fulfilling both requirements. 
Therefore, we developed a method to determine the
observational ZAHB level when the HB is well populated.  This can be
applied each time the observational HB is populated in the horizontal
part, even if there are no stars in the instability strip.

\begin{figure}
\begin{center}
  \mbox{\epsfxsize=0.9\hsize\epsffile{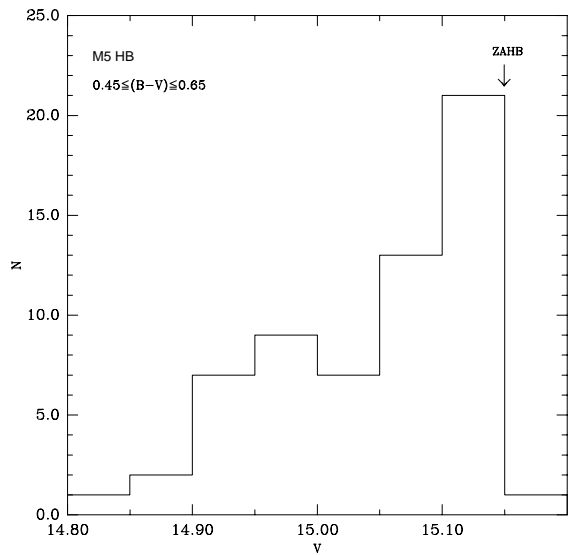}}
\end{center}
\caption{Brightness distribution of M5 HB stars in the colour region
$0.45<(B-V)<0.65$. The arrow marks the ZAHB level determined by both
the total number of stars above this bin and by the decrease in the
number of stars per bin}
\label{hbhisto}
\end{figure}

Theoretically, all observed HB stars should be at least as bright as
the ZAHB. Thus, the lower envelope (for well-populated HBs) to the
observed HB provides a reasonable estimate for the ZAHB luminosity
(Sandage 1990).  In practice, however, photometric errors, field stars
and other objects not belonging to the HB will spread out this lower
limit and a more statistical approach is necessary. To this end we
have looked into the brightness-distribution of HB stars in a few
colour bins. For each colour bin count histograms for brightness bins
were created; the brightness bins were typically 0.04-0.05 mag wide
(depending on the HB population).  Formally,
we set the ZAHB level to the upper brightness of that bin which shows
a decrease in star counts by a factor $\geq$ 2 and where the brighter
bins contain more than 90\% of all candidate HB stars under consideration. This
is illustrated in Fig.~\ref{hbhisto} 
for M5.  For all the clusters to which we have applied the vertical
method these two conditions were always
fulfilled by the same luminosity bin.  For M68 and M15, our method
reproduces the ZAHB levels at the RR~Lyrae instability strip as
determined by Eq.~\ref{vzahb} (Paper~I) within 0.01 mag, i.e.\ within
the binning error. In this paper we will adopt for M68 the ZAHB
luminosity and the associated formal error (the width of the
luminosity bin) determined with the new method;
the resulting absolute age is the same as in
Paper~I, but the formal error is lower (see Table~1).  Note that
the way we define the ZAHB level leads to a systematic overestimate of
the ZAHB brightness of order half the bin width and
consequently to ages slightly too high; the uncertainty, however,
remains below $\approx$ 0.04-0.05 mag.

To check the influence of the chemical composition on our
vertical-method age estimates, we have shown in Paper~I 
(but see also Salaris et al.\ 1994; Chaboyer 1995) 
that a variation of the metallicity by a factor of 2 or the use of an
initial helium abundance $Y$ different by 0.01 in the theoretical
isochrones 
changes the derived ages by slightly less than 1 Gyr. When taking
into account also helium and heavy element diffusion, contrary to our
expectations expressed in Paper~I, the vertical-method
ages derived decrease by less than 1 Gyr because of balancing effects
on TO and ZAHB (Castellani et al.\ 1996).

\subsection{Relative age determination}

In deriving the relative GC ages we have applied the horizontal
method, following the procedure presented by VandenBerg et al.\ (1990).
The ridge lines of two clusters are shifted horizontally in order to
match their TO colours, and then vertically to force coincidence
between the main sequences at a position 0.05 mag redder than the TO. 
Differences
in the RGB colour (fixed for example at a point 2.5 mag more luminous
than the MS reference point) correspond to age differences derived
from our new theoretical isochrones.  
As already discussed by VandenBerg et al.\ (1990), the precise point on the
RGB is of little significance, since the RGBs run essentially parallel
to one another. We have evaluated {\em mean} $\Delta(B-V)$ differences
from the cluster fiducials by considering, when possible, a
magnitude range along the RGB of typically 0.5-1.0 mag (depending on
the extension of the fiducial line), starting approximately from the
point 2.5 mag more luminous than the MS reference point. The formal
error of the relative ages derived by means of this procedure,
estimated from the uncertainty in measuring the shift in the position
of the RGB with respect to the reference cluster is about 0.5 Gyr.

As for the reliability of the age scaling derived adopting the
$\Delta(B-V)$ technique, the following points have to be mentioned:

\noindent
i) A comparison of the $\Delta(B-V)$ scaling with respect to the age
at fixed global metallicity, by adopting the oxygen-enhanced Bergbusch
\& VandenBerg (1992), the scaled-solar Straniero \& Chieffi
(1991) and our own isochrones shows a good agreement in spite of
different codes, different effective temperature normalization and/or
completely different input physics and heavy element distribution.
The relative ages obtained in the
three cases differ by no more than 20\%.

\noindent
ii) The effect of a variation in the initial helium content is almost
negligible.  A variation of $\delta Y= \pm 0.01$ induces a change in the
relative ages of only $\approx 0.15$ Gyr.  

\noindent 
iii) A variation in the global metallicity by a factor of 2 changes
the relative ages by less than 0.5 Gyr.

\noindent
iv) A global change of the mixing length parameter $\alpha_{\rm MLT}$ for each
metallicity affects {\em absolute} ages derived by means of the
horizontal method much more than the {\em relative} ones. A variation of
$\alpha_{\rm MLT}$ by 
$\pm 0.1$ changes the relative ages by 0.10-0.15 Gyr only, while changing
the absolute ages by $\approx$ 1 Gyr.

\noindent
v) The influence of the helium and heavy element diffusion on the
relative ages obtained by means of the horizontal method is almost
negligible (Cassisi 1996, private communication).  A check at $Z=0.0002$
shows that the age differences determined with and without including
diffusion agree within 10\%.

\noindent
vi) The scaling of $\Delta(B-V)$ with respect to the age is
independent of the absolute age only for ages higher than a certain
value (depending on the metallicity). Below this limit, which, for
example, is around 11-12 Gyr at $Z=0.0002$, the age
differences depend on the absolute ages assumed for the reference
clusters; it is therefore important to fix the absolute age of the
reference cluster within each group by means of the vertical method.

\section{Cluster ages -- absolute and relative}

In this section we will discuss separately the absolute and relative
age determinations of the clusters within each of the four metallicity
groups. All the selected clusters, their metallicity, HB type,
galactocentric distance ($R_{GC}$), the derived ZAHB luminosity (and the
associated error) for the reference clusters, relative and
absolute ages (with their formal errors), are displayed in Table~1.
The ZAHB luminosities refer to the instability strip for M68 and
to the red boundary of the instability strip for NGC6584, NGC3201 and M5,
which have a well developed HB but no homogeneus RR Lyrae photometry.
In the case of NGC6171 and NGC6652 the ZAHB luminosity corresponds to the red HB.
The HB types come from Chaboyer et al.\ (1996). 
An evaluation of
the galactocentric distances (in Kpc) has been obtained by 
applying the following equation: 

\begin{equation}
R_{GC}=((R_{GC}^{\odot})^2+d^2-2dR_{GC}^{\odot}cos(l)cos(b))^{0.5}
\label{galdis}
\end{equation}

where $b$ and $l$ are the cluster galactic coordinates,
$log(d)=(((m-M)_{o}+5)/5)-3$ (using $A_{V}$=3.3E(B-V)), and
the Sun's galactocentric distance is set to $R_{GC}^{\odot}$=8.0 Kpc.
The apparent distance moduli were obtained using our theoretical ZAHB
models and the ZAHB luminosities given in Table 1 for the reference
clusters, or the average HB magnitudes given by Chaboyer et al.\
(1996; in the case of NGC6366 the average HB luminosity comes from
Alonso et al.\ 1997), translated into ZAHB luminosities by means of
Eq.~\ref{vzahb}. The reddening values come from our work (and paper I
in the case of M68) for the ``reference'' clusters, from Alonso et
al.\ (1997) for NGC6366, and from Chaboyer et al.\ (1996) for all
others. 

A simple estimate of the formal error in the absolute age for the
reference clusters is obtained as in Paper~I, by statistically adding
the formal uncertainties on the ZAHB level (displayed in Table 1) and
on the TO luminosity (as derived from the original papers) in order to
obtain the error in the observed $\Delta(V)$ value; this error is then
transformed into an uncertainty in the age by using the theoretical
isochrones.  For the other clusters the formal error in the relative
ages derived by means of the horizontal method (see the previous
section) is statistically added to the error in the age of the
corresponding reference cluster.

At this point we have to comment on the meaning of the formal errors
given in Table~1. These errors have no true statistical meaning and
therefore the results of the statistical analyses of Sect.~4 should be
considered as being indicative and not rigorous. Chaboyer et al.\
(1996), who commented on this, have argued that the errors quoted by
the observers correspond to $1.63 \sigma$, and since these errors
referring to the HB or TO brightness are of a similar kind as those
given in Table~1, our errors could be considered as
corresponding to $1.63 \sigma$. However, we prefer to be more
conservative and assume that they are of the order of $1\sigma$. An
argument in favour of this is that according to
Chaboyer et al.\ (1996) the average $1\sigma$ error for all their
clusters is 0.083 mag -- corresponding to $\approx$1 Gyr, which is a similar age
uncertainty as we give in Table~1.

\begin{table*}
\caption[ ]{Halo Globular Cluster data. The columns display
respectively: cluster name, global metallicity (including
$\alpha$-enhancement), absolute age with the 
associated formal error, relative age with respect to the reference
cluster, HB type, galactocentric distance (in kpc),
estimated level of the observational ZAHB (only for reference
clusters). The absolute age is obtained by means of the vertical method for
the reference clusters and by adding the relative ages for
all other clusters in the same group.}
\begin{tabular}{lrrrrrr}
\hline\noalign{\smallskip} 
Name &$\rm [M/H]$& Age & Rel. age & HB type & $R_{GC}$ & $V_{zahb}$\\ 
\noalign{\smallskip} \hline\noalign{\smallskip} 
& & \multicolumn{3}{c}{$-2.1\leq{\rm[M/H]}<-1.6$} \\
 NGC4590~(M68)& -1.90 & 12.2$\pm$1.0 & & 0.44 & 10.2 & 15.72$\pm$0.04\\
 NGC6341~(M92)& -2.04 & 11.8$\pm$1.1 & -0.4 & 0.88 & 9.5 & \\
 NGC7099~(M30)& -2.03 & 12.7$\pm$1.1 & 0.5 & 0.88 & 7.3 & \\
 NGC7078~(M15)& -1.95 & 12.2$\pm$1.1 & 0.0 & 0.72 & 10.5 & \\ 
 NGC6397 & -1.70 & 12.2$\pm$1.1 & 0.0 & 0.93 & 5.9 & \\
 NGC2298 & -1.65 & 12.5$\pm$1.1 & 0.3 & 0.93 & 17.0 & \\
 Arp2 & -1.65 & 10.6$\pm$1.1 & -1.6 & 0.86 & 24.4 & \\
 Rup106 & -1.65 & 10.1$\pm$1.1 & -2.1 &-0.82 & 18.7 & \\ 
& & \multicolumn{3}{c}{$-1.6\leq{\rm[M/H]}<-1.3$} \\
 NGC6584 & -1.34 & 11.0$\pm$1.1 & &-0.09 & 6.6 & 16.60$\pm$0.05\\
 NGC3201 & -1.36 & 10.5$\pm$1.2 & -0.5 & 0.08 & 10.1 & 14.90$\pm$0.05\\
 NGC1904~(M79) & -1.49 & 11.0$\pm$1.2 & 0.0 & 0.89 & 19.0 & \\
 NGC5272~(M3) & -1.46 & 11.0$\pm$1.2 & 0.0 & 0.08 & 11.9 &\\
 NGC6254~(M10)& -1.40 & 11.0$\pm$1.2 & 0.0 & 0.94 & 4.7 & \\
 NGC6752 & -1.34 & 10.5$\pm$1.2 & -0.5 & 1.00 & 5.2 & \\
 NGC7492 & -1.31 & 11.0$\pm$1.2 & 0.0 & 0.90 & 24.8 & \\
& & \multicolumn{3}{c}{$-1.3\leq{\rm[M/H]}<-0.9$} \\
 NGC5904~(M5) & -1.20 & 10.9$\pm$0.8 & & 0.37 & 6.2 & 15.15$\pm$0.05\\
 Pal5 & -1.27 & 9.3$\pm$0.9 & -1.6 & -0.40 & 17.3 & \\
 NGC288 & -1.20 & 9.8$\pm$0.9 & -1.1 & 0.95 & 11.7 & \\
 NGC1851 & -1.13 & 8.9$\pm$0.9 & -2.0 &-0.33 & 17.6 & \\
 NGC362 & -1.07 & 9.7$\pm$0.9 & -1.2 &-0.87 & 9.2 & \\
 Pal12 & -0.94 & 7.5$\pm$0.9 & -4.0 &-1.00 & 17.0 & \\
& & \multicolumn{3}{c}{$-0.9\leq{\rm[M/H]}<-0.6$}\\
 NGC6171~(M107) & -0.79 & 11.0$\pm$1.1 & &-0.76 & 3.6 & 15.72$\pm$0.04\\
 NGC6652 & -0.69 & 8.0$\pm$1.2 & &-1.00 & 1.6 & 15.95$\pm$0.05\\
 Ter7 & -0.80 & 6.5$\pm$1.2 & -4.5 &-1.00 & 27.5 & \\
 NGC6366 & -0.79 & 13.2$\pm$1.2 & 2.2 &-1.00 & 4.9 & \\
 \noalign{\smallskip} \hline
\end{tabular}
\end{table*}

\subsection{Metal-poor clusters: $-2.0\leq{\rm[M/H]}<-1.6$} 

The clusters belonging to this group are M68 (Walker 1994), M15
(Durrell \& Harris 1993), M92 (Stetson \& Harris 1988), M30 (Richer
et al.\ 1988), NGC6397 (Buonanno et al.\ 
1989), NGC2298 (Janes \& Heasley 1988), Rup106 (Buonanno et al.\ 1993)
and Arp2 (Buonanno et al.\ 1995a).

The reference cluster is M68, and its age determined as in Paper~I by
means of the vertical method is 12.2$\pm$1.0 Gyr. 
The distance modulus we derive from our ZAHB models is $15.26\pm0.06$.
The relative ages of the other clusters with respect to M68 have been
derived by means of the horizontal method and are displayed in
Table~1.

Recently, Reid (1997) has used HIPPARCOS parallaxes for nearby
metal-poor subdwarfs to determine improved absolute V magnitudes, 
which we used in order to derive the distance modulus by means of the
MS fitting method (see, e.g., Sandquist et al.\ 1996). 
We have considered only objects with an error
in the parallax determination of less than 12\% and a metallicity 
including an average $\rm [\alpha/Fe]=0.3$ not higher than $Z\approx 0.0006$,
such that the colour corrections to be applied to this
empirical subdwarf sequence to match the metallicity of M68 are minimized.
In order to have stars representative of the unevolved part of the MS,
only objects fainter than absolute magnitude 5.8 have been selected.
Six stars were found which satisfy all these requirements. Assuming the
M68 metallicity given in Tab.~1 with an error of $\pm0.20$ dex, a
reddening of $0.07\pm0.01$ as given by Walker (1992b), and the colour
corrections given by our isochrones, which we extended for this
purpose down to lower masses, we obtain $(m-M)_{V,subdw}$=15.43$\pm$0.19. 
Within the errors this value, obtained by using the subdwarfs with
HIPPARCOS parallaxes, agrees with that obtained from the ZAHB.

Gratton et al.\ (1997) performed a similar study based on high precision
trigonometric parallaxes from HIPPARCOS, coupled with accurate high resolution
spectroscopic determinations of $\rm [Fe/H]$ and $\rm [\alpha/Fe]$ for a sample
of about 100 subdwarfs. The average $\alpha$-element enhancement that
they determine is about 0.3 dex for ${\rm [Fe/H]}<-0.5$.
With these data they define the absolute
location of the empirical MS as a function of $\rm [M/H]$, and determine the
distances to 9 GCs by means of the MS fitting method, using a relation
for the scaling of the $(B-V)$-colour of the empirical MS with respect
to the metallicity that matches the observations and is 
in good agreement with the one derived from our models.
Among their sample of GCs there are two of our template clusters,
namely M68 and M5 (see below). For M68 they get
$(m-M)_{V,subdw}=15.37\pm0.10$ to be compared with our value of
15.26, which becomes $(m-M)_{V}=15.23\pm0.06$, if we take into account
the slightly higher metallicities used by Gratton et al.\
(1997). Thus, also this distance modulus derived from HIPPARCOS
subdwarfs agrees within the errors with our value.

\begin{figure}
\begin{center}
  \mbox{\epsfxsize=0.9\hsize\epsffile{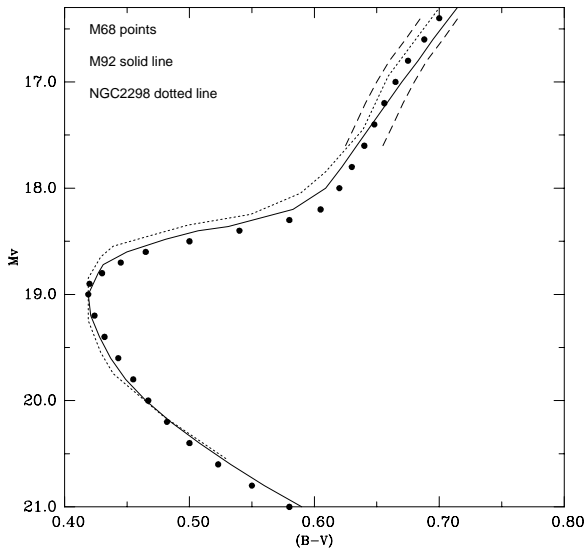}}
\end{center}
\caption{Comparison of CMD ridge lines registered to that of M68 as
explained in the 
text. The dashed lines on both sides of the M68 RGB indicate an age
difference of $\pm 1$ Gyr with respect to this cluster}
\label{hor1group1}
\end{figure}

As an example for the determination of relative ages we show in
Fig.~\ref{hor1group1} the fiducial lines of NGC2298 
and M92 registered to that of M68 as described in the previous
section; the two dashed
lines parallel to the RGB of M68 correspond to an age variation of
$\pm1$ Gyr with respect to the age of M68.  When neglecting Rup106
and Arp2, this group is remarkably homogeneus in age (an
occurrence already discussed by VandenBerg et al.\ 1990 and Straniero \&
Chieffi 1991), the maximum age difference with respect to M68 being 
$\approx 0.5$ Gyr.

\begin{figure}
\begin{center}
  \mbox{\epsfxsize=0.9\hsize\epsffile{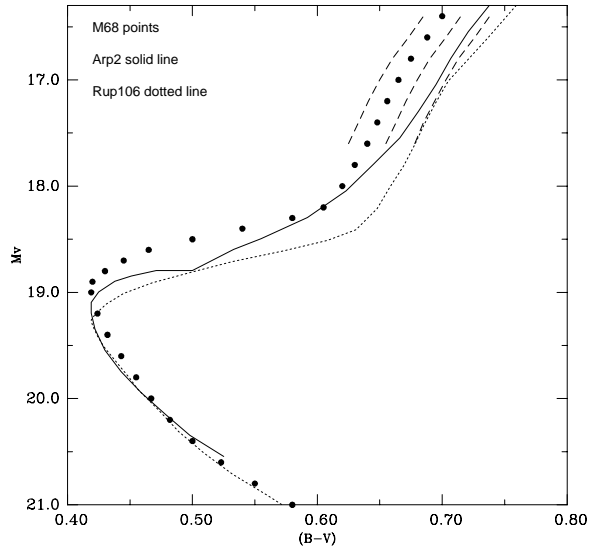}}
\end{center}
\caption{As Fig~3, but for Rup106 and Arp2; in this case the dashed
lines correspond 
to age differences of +1, -1 and -2 Gyr with respect to M68}
\label{hor2group1}
\end{figure}

Rup106 and Arp2 have been included in this group following the
metallicity determinations by Buonanno et al.\ (1993, 1995a), which
are based on
the characteristics of the observed cluster RGBs.  The final value for
[M/H] has been obtained by adding the contribution of the
$\alpha$-elements as for the other clusters.  The main lines of
Rup106, Arp2 and M68 are displayed in Fig.~\ref{hor2group1}, shifted as
before; Rup106 and Arp2 appear to be younger by
$\approx 2$ Gyr with respect to M68. This age difference is only half
as large as that claimed by Buonanno et al.\ (1993, 1995) with
respect to an 'average' metal-poor cluster, obtained by averaging the
fiducial lines of M92 (Stetson \& Harris 1988), M68 (McClure et al.\ 
1987), NGC6397 (Buonanno et al.\ 1989), M15 (Fahlman et al.\ 1985)
and M30 (Richer et al.\ 1988). 

In this work we have compared Rup106 and Arp2 with the new M68
photometry by Walker (1994), and the differences in the average
$\Delta(B-V)$ values of Rup106 and Arp2 with respect to M68 are,
respectively, +0.041 mag and +0.030 mag, almost coincident with
the values +0.041 mag and +0.028 mag found by Buonanno et al.\ (1993,
1995a). The smaller age difference is due to the fact that the
relative ages determined with the horizontal method depend on the
absolute age of the template cluster for the age range we are dealing
with (the same behaviour is found when considering, for example, the
Straniero \& Chieffi 1991 or the Bergbusch \& Vandenberg 1992 isochrones). 
If an age of 16 Gyr were assumed for M68, we would find
that Rup106 is younger than M68 by $\approx$4 Gyr, 
and that Arp2 is $\approx$1 Gyr older than Rup106, in
good agreement with
the results found by Buonanno et al (1995a).

\begin{figure}
\begin{center}
  \mbox{\epsfxsize=0.9\hsize\epsffile{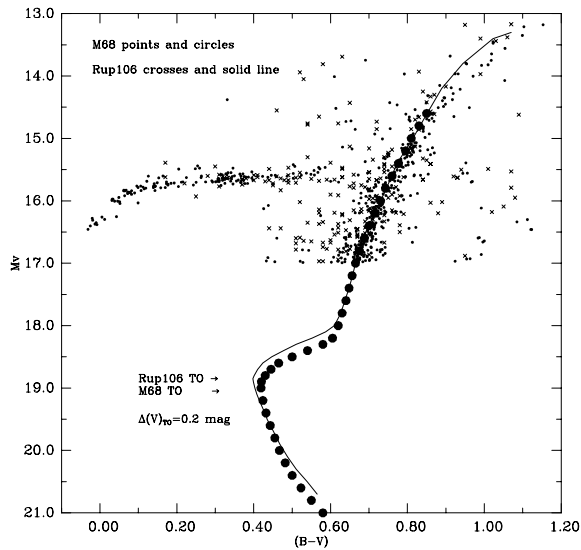}}
\end{center}
\caption{Rup106 CMD and ridge line shifted in order to superimpose RGB
and HB to the M68 ones. The cluster TO luminosities are indicated}
\label{M68Rup106}
\end{figure}

For testing in a different way the reliability of the relative ages of
Rup106 and Arp2 with respect to M68 as derived from the horizontal
method, we have checked if the same age difference is
consistent with that obtained from the vertical method (in a very similar
way as in Buonanno et al.\ 1993).  As displayed in Fig.~\ref{M68Rup106}, we have
considered the Rup106 photometry (the Arp2 photometry shows only a very
poorly populated blue HB), and we have shifted horizontally and
vertically the CMD and ridge line in order to superimpose them on the
HB and RGB of M68; note the almost coincident shapes of
the RGBs, which indicate a very similar metallicity for these two
clusters.  
The TO luminosities given by the observers are 21.05 mag
for Rup106 and 19.05 mag for M68; the vertical shift applied to Rup106
is -2.2 mag, and the horizontal one is -0.17 mag.
The difference in the TO luminosities gives the age
difference from the vertical method. Shifted in this way, the TO of
Rup106 differs by 0.2 mag (Rup106 TO being more luminous) with respect
to the M68 TO, and correspondingly Rup106 is $\approx 1.7$ Gyr younger
than M68, in very good agreement with the value derived from the
horizontal method.

\subsection{Intermediate metal-poor clusters: $-1.6\leq{\rm[M/H]} <-1.3$}

In this metallicity range we have considered the following clusters:
NGC6584 (Sarajedini \& Forrester 1995),
M3 (Ferraro et al.\ 1996),
M79 (Ferraro et al.\ 1992), NGC6752 (VandenBerg et al.\ 1990),
NGC7492 (Cot\'e et al.\ 1991), M10 (Hurley et al.\ 1989)
and NGC3201 (Covino \& Ortolani 1997).

\begin{figure}
\begin{center}
  \mbox{\epsfxsize=0.9\hsize\epsffile{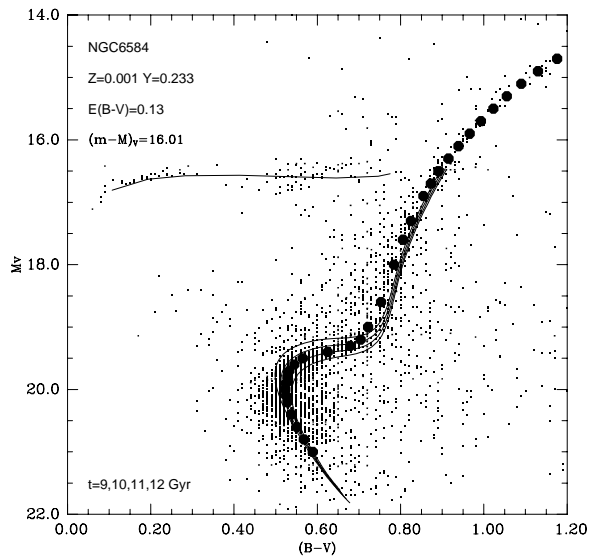}}
\end{center}
\caption{Isochrones for ages between 9 and 12 Gyr and ZAHB theoretical
models compared to the CMD and the ridge line of NGC6584 (Sarajedini
\& Forrester 1995)}
\label{N6584fit}
\end{figure}

The reference cluster is NGC6584. From the vertical method, using a
metallicity $Z=0.001$, we derive t=11.0$\pm 1.1$ Gyr (see
Fig.~\ref{N6584fit}; we obtain
an apparent distance modulus of $(m-M)_{V} =16.01\pm 0.05$ and a
reddening of $E(B-V)=0.13$, in agreement with previous estimates ranging
between 0.07 and 0.15.  By applying the horizontal method, we derive
age differences not bigger than 0.5 Gyr for all clusters in the
sample (see Table~1).

In Fig.~\ref{N6584fit} (as in Fig.~\ref{N6171fit})
the observational data appear to be quantized; this is due to the fact
that the 
available files with the photometric data provide V and (B-V) values
with only two decimal digits. However, the fiducial line and the
TO luminosity we use are the ones provided in the cited papers and were
derived by the authors using the original data with more than two
decimal digits. Moreover, this quantization does not affect the derived 
observational value of the ZAHB brightness, which is determined with an
error of typically $\pm$0.05 mag.

In the case of NGC6584
we have verified that using a metallicity of $Z=0.0006$, which
corresponds to the lower boundary of the error range associated
with the metallicity determination ($\rm [M/H]$ of 0.2 dex),
the isochrone fit can be improved.
The absolute age is changed by only $\approx 0.5$ Gyr, and
the relative ages determined by means of the horizontal
method are affected much less.

\begin{figure}
\begin{center}
  \mbox{\epsfxsize=0.9\hsize\epsffile{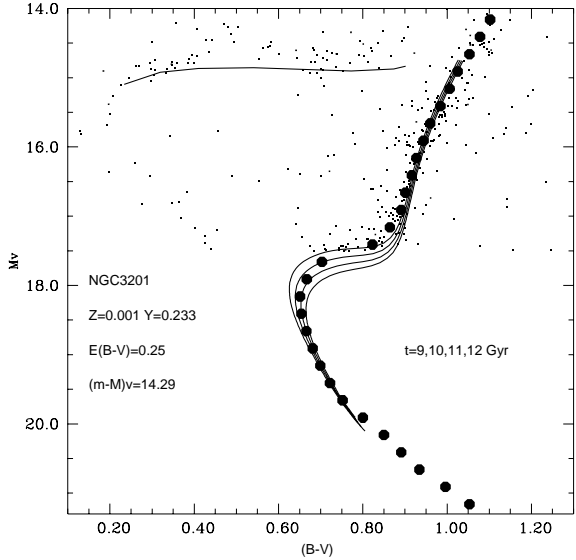}}
\end{center}
\caption{Isochrones for ages between 9 and 12 Gyr and ZAHB theoretical
models compared to the CMD and ridge line of NGC3201 (Covino \& Ortolani 1997). 
For sake of clarity, only the ridge line is displayed for the
cluster MS}
\label{N3201fit}
\end{figure}

\begin{figure}
\begin{center}
  \mbox{\epsfxsize=0.9\hsize\epsffile{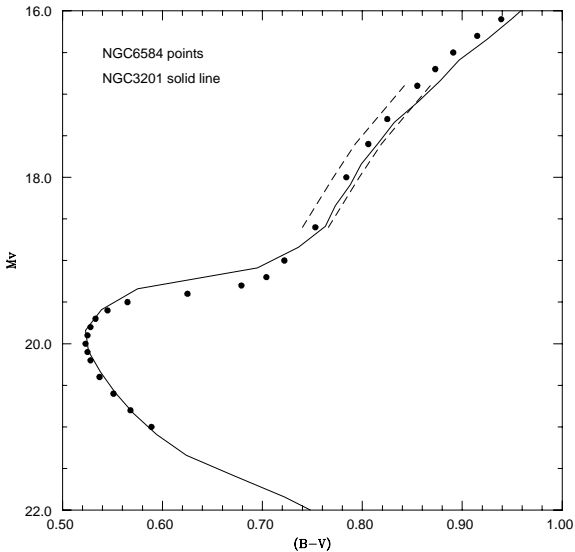}}
\end{center}
\caption{The CMD ridge line of NGC3201 registered to
that of NGC6584 as explained in the text. The dashed lines on both sides of
the NGC6584 RGB indicate an age difference of +1 and -1 Gyr with respect to NGC6584}
\label{hor1group2}
\end{figure}

We can check the consistency of the horizontal age determination with the
absolute vertical age determination for another cluster with
a well populated horizontal part of the HB, that is NGC3201 (Covino \& Ortolani 1997).
We get an age of $10.0 \pm 1.6$ Gyr from the
vertical 
method ($(m-M)_{V}=14.28\pm 0.05$, $E(B-V)=0.25$) adopting $Z=0.001$
(see Fig.~\ref{N3201fit}); this value is in very good agreement with the
age  obtained by means of the horizontal method relative to NGC6584
(see Table~1 and Fig.~\ref{hor1group2}).
  
We have also calculated the age of Rup106 in a third way
by assigning it to the
intermediate metal-poor group and by deriving its relative age with
respect to NGC6584.  These two clusters differ by $\approx 0.3$ dex in
$\rm[M/H]$, and in principle -- following the criteria adopted in this work
-- could belong to the same metallicity group. Rup106 is measured to be
1.3 Gyr younger than NGC6584; thus
the derived absolute age in this case is $9.7\pm 1.2$ Gyr, 
consistent with the value derived in the preceding subsection (see
Tab.~1).

\subsection{Intermediate metal-rich clusters:
$-1.3\leq{\rm[M/H]}<-0.9$}

This group includes M5 (Sandquist et al.\ 1996), NGC1851 (Walker 1992b),
NGC288 (Bolte 1992), NGC362 (VandenBerg et al.\ 1990), Pal12 (Stetson et
al.\ 1989) and Pal5 (Smith et al.\ 1986).

\begin{figure}
\begin{center}
  \mbox{\epsfxsize=0.9\hsize\epsffile{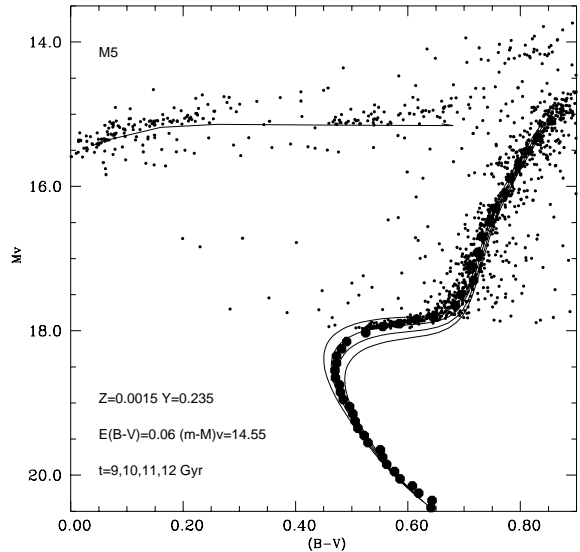}}
\end{center}
\caption{Isochrones for ages between 9 and 12 Gyr and ZAHB theoretical
models compared to the CMD and ridge line of M5 (Sandquist et al.\ 
1996). Only the ridge line is displayed for the
cluster MS, and along the RGB, at luminosities lower than the HB, 
only a subsample of stars is shown}
\label{M5fit}
\end{figure}

The reference cluster is M5. By applying the vertical method to the
recent photometric data by Sandquist et al.\ (1996), and using Z=0.0015
we get an age of
$10.9\pm 0.8$ Gyr (see Fig.~\ref{M5fit}) and $E(B-V)=0.06$,
$(m-M)_{V}=14.55\pm0.05$. If we compute the
real distance modulus, by adopting $A_{V}=3.3E(B-V)$, we get
$(m-M)_{0} =14.35 \pm 0.05$, a value that agrees well with the
value of $14.37 \pm 0.18$ found by Storm et al.\ (1994) from the
Baade-Wesselink method for two cluster RR Lyrae stars. 
Our $(m-M)_{V}$
is also confirmed by the recent result of Gratton et al.\ (1997),
$(m-M)_{V,subdw}=14.58\pm0.04$, which is based on HIPPARCOS subdwarf
parallaxes (see Sect.~3.1), even if we take into account the slightly
higher metallicity for M5 used by Gratton et al.\ (1997). In this
case, our distance modulus becomes $(m-M)_{V}=14.50\pm0.07$.

\begin{figure}
\begin{center}
  \mbox{\epsfxsize=0.9\hsize\epsffile{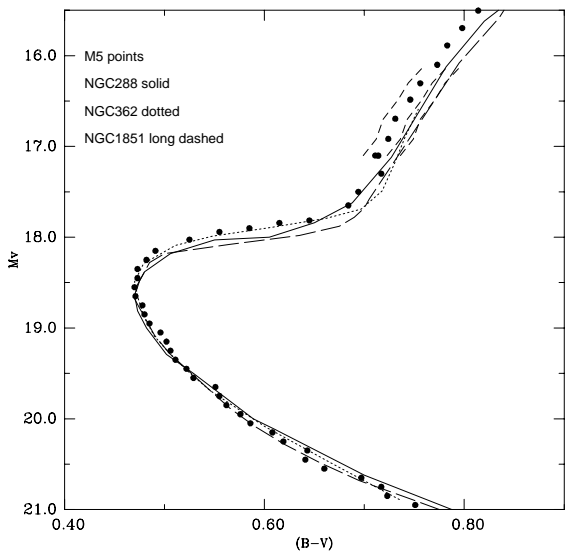}}
\end{center}
\caption{Comparison of CMD ridge lines. The dashed lines on both sides
of the M5 RGB indicate an age 
difference of +1, -1 and -2 Gyr with respect to the reference  cluster
M5}
\label{hor1group3}
\end{figure}

The relative ages with respect to M5 are displayed in Table 1. All the
other clusters are found to be significantly younger, the youngest one
being Pal12.  In particular, 
when taking into account the recent photometric study by Bolte (1992)
of NGC288, we find that NGC288 and NGC362 are practically coeval
(see Fig.~\ref{hor1group3}). This confirms the qualitative result by
Stetson et al.\ 
(1996), who found  using essentially the vertical method that
NGC1851, NGC362 and NGC288 should have the same age, thus giving
very strong evidence against age as the second parameter. Our
result is in agreement with their investigation; by using the
horizontal method, the three clusters are found to be coeval within less
than 1 Gyr (Fig.~\ref{hor1group3})).

\subsection{Metal-rich clusters: $-0.9 \leq {\rm [M/H]}<-0.6$}

\begin{figure}
\begin{center}
  \mbox{\epsfxsize=0.9\hsize\epsffile{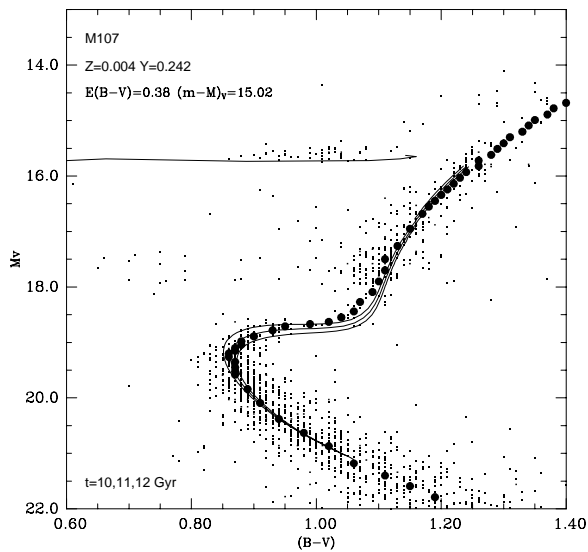}}
\end{center}
\caption{Isochrones for ages between 10 and 12 Gyr and ZAHB
theoretical models compared to the CMD and ridge line of M107
(Ferraro et al.\ 1991)}
\label{N6171fit}
\end{figure}

M107 (Ferraro et al.\ 1991), NGC6652 (Ortolani et al.\ 1994),
NGC6366 (Alonso et al.\ 1997) and Ter7
(Buonanno et al.\ 1995b) are the four clusters considered in this
group.  The reference cluster is NGC6171 (=M107). By employing isochrones for
$Z=0.004$ we get from the vertical method an age of $11.0 \pm 1.1$ Gyr,
together with $E(B-V)=0.38$ and $(m-M)_{V} =15.02 \pm 0.04$ (see
Fig.~\ref{N6171fit}). The reddening we derive agrees with previous estimates,
ranging between 0.30 and 0.48.

\begin{figure}
\begin{center}
  \mbox{\epsfxsize=0.9\hsize\epsffile{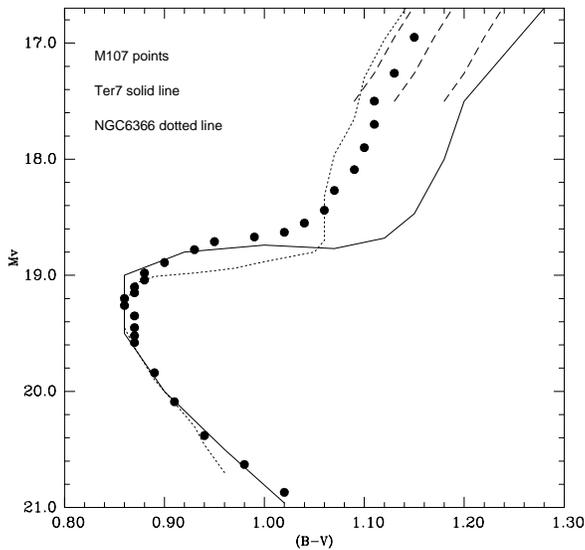}}
\end{center}
\caption{Comparison of CMD ridge lines. 
The dashed lines on both sides of the M107 RGB indicate an
age difference of +2, -2 and -4 Gyr with respect to this cluster}
\label{hor1group4}
\end{figure}

By applying the horizontal method with
respect to M107 (see Fig.~\ref{hor1group4}), we find a large age 
spread within this group: NGC6366 is $\approx 2$ Gyr older than
M107, while Ter7 is $\approx 4.5$ Gyr younger than M107. This is
the largest age spread among all the clusters considered in this
study. However some caution is required when
considering NGC6366, since  this cluster (see Harris 1993 and Alonso et al.\ 1997) is
affected by differential reddening; the determination of its relative age with respect
to NGC6171 could therefore be less accurate even if the cluster
appears undoubtedly to be an ``old'' halo GC (see also the discussion
in Alonso et al.\ 1997). 

\begin{figure}
\begin{center}
  \mbox{\epsfxsize=0.9\hsize\epsffile{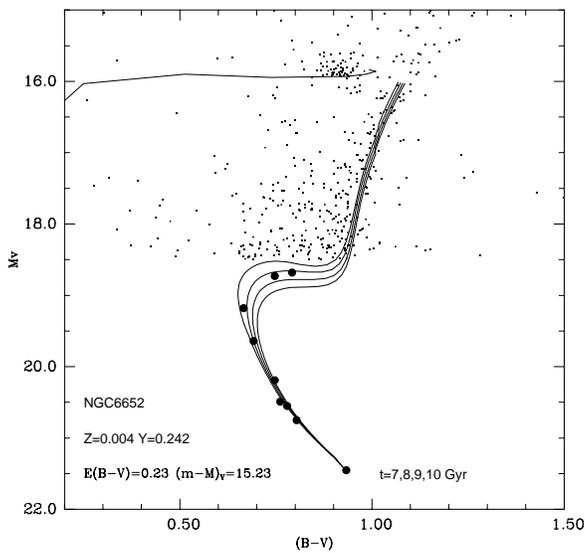}}
\end{center}
\caption{Isochrones for ages between 7 and 10 Gyr
and ZAHB theoretical models compared to the CMD and ridge line of
NGC6652 (Ortolani et al.\ 1994). Only the ridge line is displayed for the
cluster MS}
\label{N6652fit}
\end{figure}

The cluster ridge line for NGC6652 (not provided in the paper by 
Ortolani et al.\ 1994) has been derived by determining the 
median of the colour distribution within brightness bins. The
resulting TO luminosity is in very good agreement 
with the value $V_{TO}$=19.20$\pm$0.15 given by Ortolani et al.\ (1994).
Since the cluster RGB shows a large dispersion in colour and the ridge line for this
CMD region is not well defined, we have considered the ridge line only
up to the subgiant branch. Together with a very
well defined and populated red HB, this is sufficient for directly
estimating the absolute cluster age (as given in Table 1)
by means of the vertical method. By using isochrones for $Z=0.004$ we get
an age of $8.0\pm1.2$ Gyr, $E(B-V)=0.23$, $(m-M)_{V}=15.23$
(see Fig.~\ref{N6652fit}).

\begin{figure}
\begin{center}
  \mbox{\epsfxsize=0.9\hsize\epsffile{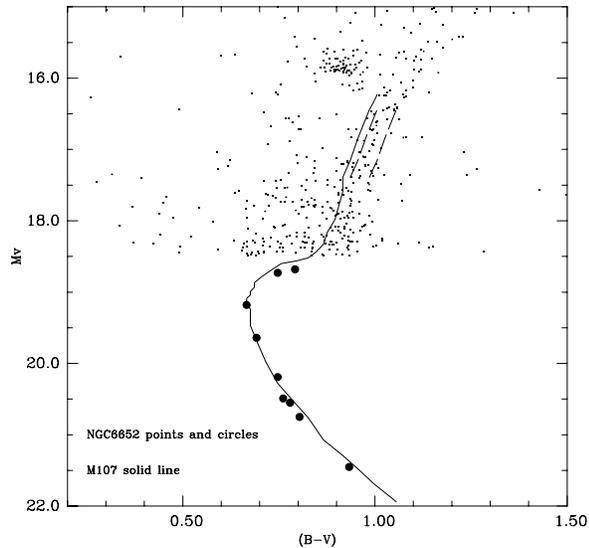}}
\end{center}
\caption{Relative age of NGC6652 with respect to M107 derived
from the horizontal method. 
The dashed lines on the right side of the M107 RGB indicate an
age difference of -2 and -4 Gyr with respect to this cluster}
\label{hor2group4}
\end{figure}

The consistency between the NGC6652 absolute age and
its relative age with respect to M107 derived by means of the horizontal method
can be checked qualitatively by registering the M107
ridge line to that of NGC6652, as shown in Fig.~\ref{hor2group4}.
Since the RGB of NGC6652 shows a large dispersion in colour, it is
not possible to derive a reliable independent estimate of its relative age with respect
to M107, but we can verify the consistency of the relative
RGB positions with the absolute ages. 
If we consider the magnitude range $M_{V}\approx$ 16-17, where the
NGC6652 RGB appears to be better defined, the M107 fiducial line
lies at the left boundary of the NGC6652 RGB,  
while the line corresponding to an age difference of -4 Gyr with
respect to M107 lies to the right of the RGB.
Recalling that the absolute ages of M107 and NGC6652 as
obtained by means of the vertical method are, respectively, 11.0 and
8.0 Gyr, the relative
positions of the RGBs are in qualitative agreement
with the difference in absolute ages.

\subsection{Comparison with previous results} 

The GC ages displayed in Table~1 can be directly
compared with the results from the work by Richer et al.\ (1996).  Their
approach has already been discussed in Sect.~1. They also arranged
the clusters into four groups, according to the cluster $\rm
[Fe/H]$-content. The
metallicity range of each group is very similar to our choice, as are the
metallicities adopted for each cluster. Only in the case of Rup106 and
Arp2 they are substantially higher. 
As for the absolute ages, we find that our
values are systematically lower by $\approx 4$ Gyr, due basically to the more
up-to-date stellar models we used (see Paper~I).

The relative ages among the metal-poor GCs in common
with Richer et al.\ (1996) are in agreement with their results within the
formal errors associated with the determination of the relative ages
($\approx 0.5$ Gyr for both this paper and Richer et al.\ 1996).

When considering the second group (intermediate metal-poor clusters),
we have in common with Richer et al.\ (1996) M3, NGC6752, NGC1904 and
NGC7492, 
which are coeval within 0.5 Gyr.  Richer et al.\ (1996) find the same
result for the first 4 cluster, while they obtain an age
1.7 Gyr higher for NGC7492. This is surprising, since we are using the same
source of photometric data for example for NGC6752 and NGC7492.  The
reason for this difference could be due to a point on the ridge line
of NGC7492 that is clearly discrepant. It is about 2.5 mag
brighter than the point on the main sequence used for registering all
clusters to the reference one. If only this point is considered for
determining the relative age, one indeed finds that
NGC7492 is around 1.7 Gyr older than NGC6752.

In the intermediate metal-poor group Richer et al.\ (1996) also
include Arp2 and Rup106.  They find that Rup106 is younger by 1 Gyr
with respect to Arp2, and by 4 Gyr with respect for example to
NGC6752. We also determined the relative age of Rup106 with respect to
clusters of the second group (see Sect.~3.4), and the result is
that it is younger by only $\approx 1.3$ Gyr. As
previously discussed, this result is due to the dependence of the
relative ages obtained by the horizontal method on the absolute age of
the reference cluster, for ages lower than a certain value depending
on the assumed metallicity.

The intermediate metal-rich clusters in common with Richer et al.\ 
(1996) are M5, NGC288, NGC362, NGC1851 and Pal12. The substantial
difference with their work is that we find NGC1851, NGC288 and NGC362
to be coeval within 1 Gyr (in agreement with the result by Stetson et
al.\ 1996 obtained by the vertical method). This seems to be 
due to the use of the new Bolte (1992) photometry for
NGC288. Had we used the NGC288 ridge line from Buonanno et al.\ (1989),
we would have obtained an age difference by $\approx$2 Gyr 
with respect to NGC362 (NGC288 being older), in agreement with the results 
by Richer et al.\
(1996). Another difference is that we have adopted the very recent M5
photometry by Sandquist et al.\ (1996), which also displays a
well-populated HB, and we find that M5 is older than NGC362 and NGC1851, while
Richer et al.\ find these three clusters to be coeval; this difference again
is due to the different data used. If we use the old data for M5 by
Richer \& Fahlman (1987, as in Richer et al.\ 1996), the results again agree
with Richer et al.\ (1996).

Among the metal-rich clusters there is only Ter7 in common with the
work by Richer et al.\ (1996). They consider 3 other clusters belonging
to the disk GC system, for which there are indications that the
original helium content could be substantially higher than $Y=0.23$ (see
Alonso et al.\ 1997 and references therein).

\section{Discussion}

The results displayed in Table 1 can be used for checking for 
the existence of an age spread and an age-metallicity relation
for halo clusters, as well as for 
testing the hypothesis that age is the so-called ``second parameter'', 
responsible for the HB morphology of galactic GCs.

Recent work by Chaboyer et
al.\ (1996) reaches the conclusion that age is the second parameter,
but the analyses by Richer et al.\ (1996) and Stetson et al.\ (1996)
do not confirm this. Checking Table 1, it is evident that the cluster pairs
Rup106--Arp2 and NGC288--NGC362 have almost the same metallicity
and ages, but completely different HB morphologies. We support
therefore the conclusion that HB morphology must, at least in part, 
be due to causes other than only metallicity and age.

\begin{figure}
\begin{center}
  \mbox{\epsfxsize=0.9\hsize\epsffile{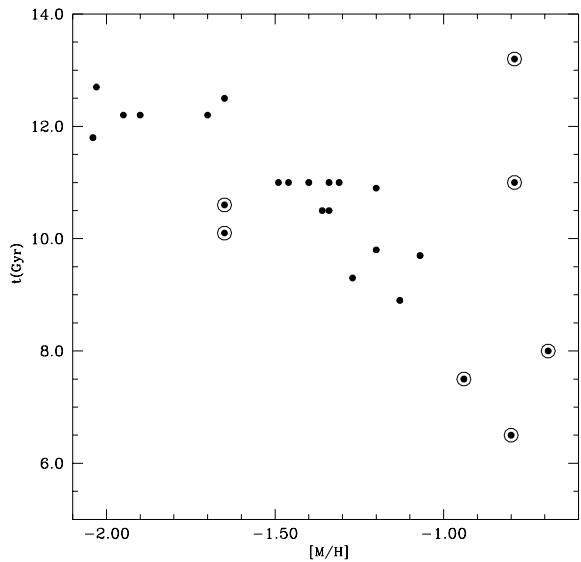}}
\end{center}
\caption{Age (in Gyr) of the 25 clusters in our GCs sample as a
function of their $\rm [M/H]$. The clusters with the circled dots
(Rup106, Arp2, Ter7, Pal12, NGC6366, M107, NGC6652) are
those excluded from the analysis is some cases (see text). 
The error on the individual ages (of order $\pm 1$ Gyr) can be found
in Table 1, while the error 
on  $\rm [M/H]$ is typically of the order of 0.20 dex}
\label{tmetal}
\end{figure}

\begin{figure}
\begin{center}
  \mbox{\epsfxsize=0.9\hsize\epsffile{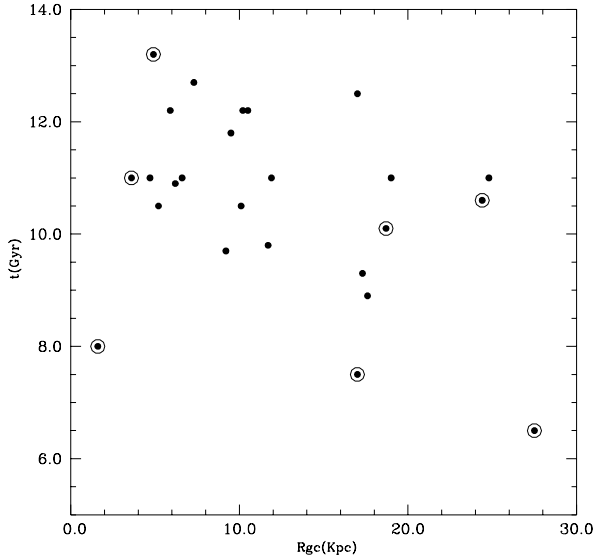}}
\end{center}
\caption{As in the previous figure, but in this case the age is
displayed as a function of the cluster galactocentric distance (in
kpc)} 
\label{tdist}
\end{figure}

As for the cluster age distribution, we find that 
if we take into account all 25 GCs considered in our investigation (see Table~1, 
Fig.~\ref{tmetal}, Fig.~\ref{tdist}), 
we obtain an average age of $\langle{\rm t}\rangle=10.6$ Gyr, with a
standard deviation $\sigma$=1.7 Gyr. 
However, as discussed in the previous section, the 
relative age of NGC6366 with respect to M107
is uncertain, due to  differential reddening,
and therefore we will not consider NGC6366 in the following analysis.
Without NGC6366 (24 clusters), 
$\langle{\rm t}\rangle=10.5$ Gyr and $\sigma$=1.6 Gyr, very close to
the results for the complete sample.
For the first three metallicity groups the average ages and the dispersions are, 
in the order of increasing metallicity, $\langle{\rm t}\rangle=11.8,\,10.9,\,9.4$
Gyr and $\sigma=0.9,\,0.2,\,1.1$
Gyr. In the most metal-rich group we have only three clusters, which have 
$\langle$t$\rangle$=8.5 Gyr. The rather large variance for the
low-metallicity group results from the two clusters Arp2 and Rup106.
If we omit these clusters (see below), we obtain $\langle{\rm t}
\rangle=12.3$ and $\sigma=0.3$.

To quantify how much of the age range among the clusters could be due
to errors and whether a real
intrinsic age range exists, we have performed the same statistical test used by 
Chaboyer et al.\ (1996). We have calculated an ``expected''
distribution for the assumption of no intrinsic age range by 
randomly generating 10000 ages using a Gaussian distribution. The mean
of the distribution was 
given by the mean age of the clusters (10.5 Gyr) , and the $\sigma$ by
the error on the individual age determinations. This is
repeated for all clusters considered, so that the final
distribution 
contains 240000 points. The F-test (Press et al.\ 1992) was then
applied in order to determine if this ``expected''
distribution has the same variance as the age distribution obtained in
our analysis.
As in Chaboyer et al.\ (1996) we state that an age range exists if
the probability that the two distributions have the same variance is
smaller than 5\%.

The F-test rejects the possibility that
the clusters are coeval with a confidence level higher than 99.8\%.
The size of the true age range ($\sigma_{range}$) can be estimated according to
$\sigma_{range}=(\sigma_{obs}^2-\sigma_{exp}^2)^{0.5}$, where $\sigma_{obs}$ is the sigma
of the actual data, and $\sigma_{exp}$ is the sigma of the expected distribution
(Chaboyer et al.\ 1996); we obtain $\sigma_{range}$=1.2 Gyr.

As for an age-metallicity relation, a formal linear fit to the data yields

\begin{equation}
t=(-3.27\pm0.53){\rm [M/H]}+(5.94\pm0.77).
\label{tm1}
\end{equation}

The linear Pearson correlation coefficient is -0.80,
implying that the confidence level for a linear correlation between
age and $\rm [M/H]$ is not high.  
The correlation coefficient for the relation between age and $R_{\rm
GC}$ is even lower (-0.34). A visual inspection of Fig.~\ref{tdist}
confirms that there is only an indication
that the youngest clusters, with the exception of NGC6652,
seem to be located in the outer Halo.

To summarize, when considering all the clusters in our sample (except
NGC6366), we find an age spread  
among the GCs ($\sigma_{range}$=1.2 Gyr),
but no statistically compelling evidence for either an age-metallicity 
or  an age-$R_{GC}$ relation. There is however an indication
that the more metal-poor clusters
are on average older then the clusters of higher metallicities, and
that the age spread within each metallicity bin tends to be higher for
increasing metallicities. Our analysis based on new stellar models
therefore confirms the results of Richer et al.\ (1996) in this
respect. 

Recently, Lin \& Richer (1992) and Buonanno et al.\ (1994) have
suggested that Pal12, Rup106, Arp2 and Ter7 (which appear to be
younger than other clusters of approximately the same metallicity)
could have been captured by the Milky Way from a companion galaxy, and
therefore represent later infall events. In this case they could not
be indicative of the halo formation phase.  This argument is based
mainly on the fact that these four clusters appear to lie along a
single great circle passing through the northern tip of the Magellanic
Stream, thus suggesting a common orbit that could be the result of an
accretion event from a companion galaxy. Although it is clear that
proper motion studies are necessary for a definitive answer, the
argument is reinforced by the fact that Ter7 and Arp2 lie very close
to the Sagittarius Dwarf Galaxy (but see also the discussion in
Chaboyer et al.\ 1996), which is currently being tidally disrupted and
absorbed by the Milky Way.

If this is the case, the previous statistical analysis should be performed
excluding these 4 objects from our halo GCs sample. Furthermore, from
Fig.~\ref{tmetal} it is evident that there are only two clusters of the
metal-rich group left -- M107 and NGC6652 (at respectively 11 and 8 Gyr). In the
following we will give results in brackets for the case when
M107 and NGC6652 are neither taken into account, and therefore the highest
metallicity clusters are disregarded completely.
The average age of the remaining 20 (18) clusters is then
$\langle {\rm t }\rangle=10.9$ (11.0) Gyr ($\sigma$=1.2 (1.1) Gyr), which is
slightly older, but has a much narrower distribution than the complete
sample. The correlation coefficient between age and metallicity is almost
-0.82 (-0.86), which is the same as for the 24 clusters.
The F-test reveals that the ``coeval'' hypothesis can be rejected with
a lower confidence level of $\approx$70\% (25\%);
the derived formal $\sigma_{range}$ is equal to 0.5 (0.2) Gyr, less than
half of $\sigma_{range}$ for the complete sample.
We therefore conclude that, once Pal12, Rup106, Arp2 and Ter7 are excluded,
the genuine halo clusters formed within $\approx 1.5$ Gyr of each
other. Our smallest sample is therefore coeval and the one including
M107 and NGC6652 cannot be excluded to be so as well. 
The formal linear regression to this sample gives 
$t=(-2.71\pm0.45){\rm[M/H]}+(7.03\pm0.66)$.

\section{Summary}

Our results can be summarized as follows:

\begin{enumerate}
\item We have determined the ages of a sample of 25 halo clusters by use of
stellar models which take into account all recent improvements
in stellar input physics data.
\item The method for obtaining ages is a combination of the
$\Delta(V)$-method for a few ``reference'' clusters and the $\Delta(B-V)$-method
for other clusters of a similar metallicity. The clusters are split
into four groups according to simila metallicity.
\item Our results, summarized in Table~1, confirm that GCs are
$\approx 12\pm 1$ Gyr old or younger, as we already claimed in
Paper~I for a subset of three metal-poor clusters. The lower ages as
compared to previous investigations are due to our new stellar physics
input and our purely theoretical approach for the HB luminosities.
\item Since age differences depend on absolute age, we obtain smaller
age differences for a given $\Delta(B-V)$. Therefore our sample
becomes more homogeneous even if we use the same original data as in
previous papers. This applies, e.g., for Rup106 and Arp2 with respect
to M68. 
\item Several cross checks (e.g.\ for Rup106) result in consistent
ages.
\item NGC6366 is the oldest cluster of our sample with $13.2\pm1.2$
Gyr, but the photometry for this cluster might be affected by
differential reddening, so we excluded it from the analysis in
Sect.~4. The next oldest cluster is M30 with $12.7\pm1.1$. 
\item We confirm earlier results that the metal-poor clusters
form a very coeval group and that the more metal-rich groups show a
larger age spread.
\item For the whole sample, which has not been selected on any
specific grounds except that the photometric quality should allow the
application of our age determination methods, we obtain a mean age of
$10.6\pm 1.7$ Gyr ($1\sigma$-error) and reject the assumption that all
clusters are coeval. A linear correlation between metallicity and age
is not confirmed.
\item For samples with all ``peculiar'' and  metal-rich clusters
excluded, the mean age becomes better defined ($11.0\pm1.1$ Gyr). 
The age range of a large sample of clusters with the same average age
and individual errors as ours would be only 0.2 Gyr (1$\sigma$
range). The clusters in this sample are coeval. This is
still true, if we include M107 and NGC6652, although the probability
for this hypothesis is lower.
Whether or not the assumption of a common age can be rejected safely,
depends critically on the inclusion of individual clusters. 
Clearly, the sample size is too small for any reliable
conclusions. 
\item There is no evidence for any correlation between age and
galactocentric distance. 
\item Known counter-examples against the hypothesis that age is the second
parameter affecting HB morphology are confirmed. NGC288 and NGC362
have the same age. This result does not agree with Richer et al.\
(1996), and is due to new photometric data of NGC288, but it is in agreement
with the results by Stetson et al.\ (1996).
\item Other differences with respect to Richer et al.\ (1996) can be
explained in terms of our new models, lower absolute ages, or different
original data.
\end{enumerate}

We confirm and substantiate the results of Richer et al.\
(1996) in large parts, although differences for individual clusters
exist, and we determine significantly lower ages for all clusters. 
According to both investigations the more metal-poor GCs all formed
within 1 Gyr and throughout the whole halo. The more metal-rich
systems possibly formed another Gyr later and over a somewhat longer
timescale. The cluster population is contaminated by a few clusters
not fitting into this simple picture. Consistent and high-quality
photometric data for a large sample of clusters is needed to confirm
our results.

\begin{acknowledgements}
We are grateful to Drs.~Alexander and Rogers for computing
special opacity tables for our purposes, E.L. Sandquist for providing us
with his excellent M5 photometry before publication. S. Covino and S. Ortolani 
are acknowledged for providing us with their NGC3201 photometry.
It is a pleasure
to thank M.~Bartelmann, J.~Guerrero and A.~Piersimoni for helpful
discussions and D.~Syer for polishing our English. 
\end{acknowledgements}


\begin{thebibliography}{99}
\bibitem{ref.1} Alonso A., Salaris M., Martinez-Roger C., Straniero O.,
Arribas S., 1997, \aap\, in press
\bibitem{ref.2} Alexander D.R., Ferguson J.W., 1994, \apj\ 437,879
\bibitem{ref.3} Bergbusch P.A., VandenBerg D.A., 1992, \apjs\  81, 163
\bibitem{ref.4} Bolte M., 1992, \apjs\ 82, 145
\bibitem{ref.5} Buonanno R., Corsi C.E., Fusi Pecci 1989, \aap\ 216, 80 
\bibitem{ref.6} Buonanno R., Corsi C.E., Fusi Pecci F., Richer H.B., Fahlman G.G., 1993,
\aj\ 105, 184 
\bibitem{ref.7} Buonanno R., Corsi C.E., Fusi Pecci F., Richer H.B., Fahlman G.G., 1994,
\apj\ 430, L121 
\bibitem{ref.8} Buonanno R., Corsi C.E., Fusi Pecci F., Richer H.B., Fahlman G.G., 1995a,
\aj\ 109, 650  
\bibitem{ref.9} Buonanno R., Corsi C.E., Pulone L.,
Fusi Pecci F., Richer H.B., Fahlman G.G., 1995b,
\aj\ 109, 663 
\bibitem{ref.10} Buser R., Kurucz R.L., 1978, \aap\ 70, 555 
\bibitem{ref.11} Buser R., Kurucz, R.L., 1992, \aap\ 264, 557 
\bibitem{ref.12} Carretta E., Gratton R.G., 1997, \aas\ 121, 95
\bibitem{ref.13} Carney B.W., Storm J., Jones R.V., 1992,  \apj\ 386, 663 
\bibitem{ref.14} Castellani V., Ciacio F., Degl'Innocenti S., Fiorentini G.,
1996, \aap\ in press 
\bibitem{ref.15} Chaboyer B., 1995, \apj\ 444, L9 
\bibitem{ref.16} Chaboyer B., Kim Y.-C., 1995, \apj\ 454, 767
\bibitem{ref.17} Chaboyer B., Demarque P., Sarajedini A., 1996, \apj\
459, 558
\bibitem{ref.18} Chaboyer B., Sarajedini A., Demarque P., 1992, \apj\ 
394, 515
\bibitem{ref.19} Chieffi A., Straniero O., 1989, \apjs\ 71, 47
\bibitem{ref.60} Clementini G., Carretta E., Gratton R., Merighi R.,
Mould J.R., McCarthy J.K., 1995, \aj\ 110, 2319
\bibitem{ref.20} Cot\'e P., Richer H.B., Fahlman G.G., 1991, \aj\ 102, 1358
\bibitem{ref.61} Covino S., Ortolani S., 1997, \aap\ 318, 40
\bibitem{ref.21} D'Antona F., Caloi V., Mazzitelli I., 1997, \apj\ 477, 519
\bibitem{ref.62} de Boer K.S., Tucholke H.-J, Schmidt J.H.K., 1997, \aap\ 317, L23
\bibitem{ref.63} Dorman B., 1992, \apjs\ 81, 221
\bibitem{ref.22} Durrel P.R., Harris W.E., 1993, \aj\ 105, 1420 
\bibitem{ref.64} Feast M.W., Catchpole R.M., 1997, \mn\ 286, L1
\bibitem{ref.23} Ferraro F.R., Clementini G., Fusi Pecci F., Buonanno, R., 1991, \mn\ 252, 357 
\bibitem{ref.24} Ferraro F.R., Clementini G., Fusi Pecci F., 
Sortino R., Buonanno, R. 1996, \mn\ 256, 391 
\bibitem{ref.25} Ferraro F.R. et al.\ 1996, \aas\ submitted 
\bibitem{} Gratton R.G., Fusi Pecci F., Carretta E., Clementini G.,
Corsi C.E., Lattanzi M. 1997, \apj\ submitted
\bibitem{ref.26} Harris H.C., 1993, \aj\ 106, 604
\bibitem{ref.27} Hurley D.J.C., Richer H.B., Fahlman, G.G., 1989, \aj\ 98, 2124
\bibitem{ref.28} Iglesias C.A., Rogers F.J., 1996, \apj\ 464, 943
\bibitem{ref.29} Lin D.N.C., Richer H.B., 1992, \apj\ 388, L57
\bibitem{ref.30} Mazzitelli I., D'Antona F., Caloi, V., 1995, \aap\ 302, 382
\bibitem{ref.31} McClure R.D., VandenBerg D.A., Bell R.A., Hesser J.E.,
Stetson P.B., 1987, \aj\ 93, 1144
\bibitem{ref.59} Ortolani S., Bica E., Barbuy B., 1994, \aap\ 286, 444
\bibitem{ref.32} Press W.H., Teukolsky S.A., Vetterling W.T., Flannery B.P., 1992,
Numerical Recipes (2d ed.; Cambridge: Cambridge Univ. Press) 
\bibitem{} Reid I.N. 1997, \aj\ in press
\bibitem{ref.33} Richer H.B., Fahlman G.G., 1987, \apj\ 316, 189
\bibitem{ref.34} Richer H.B., Fahlman G.G., VandenBerg D.A., 1988, \apj\ 329, 187
\bibitem{ref.35} Richer H.B., et al., 1996, \apj\ 463, 602
\bibitem{ref.36} Rogers F.J., Iglesias C.A., 1992, \apjs\ 79, 507
\bibitem{ref.37} Rogers F.J., Swenson F.J.,  Iglesias C.A., 1996, \apj\ 456, 902
\bibitem{ref.38} Salaris M., Chieffi A., Straniero O., 1993, \apj\ 414, 580
\bibitem{ref.39} Salaris M., Degl'Innocenti, S., Weiss, A.,  1997,
\apj\ 479, 665
\bibitem{ref.40} Salaris M., Straniero O., Chieffi A., 1994, MemSAIt
65, 693
\bibitem{ref.41} Sandage A., 1990, \apj\ 350, 603
\bibitem{ref.42} Sandage A., Cacciari C., 1990, \apj\ 350, 645
\bibitem{ref.43} Sandquist E.L., Bolte M., Stetson P.B., Hesser J.E., 1996, \apj\ 470, 910
\bibitem{ref.44} Sarajedini A., Forrester W.L., 1995, \aj\ 109, 1112
\bibitem{ref.45} Smith G.H., McClure R.D., Stetson, P.B., Hesser J.E., Bell R.A., 1986,
\aj\ 91, 842
\bibitem{ref.46} Stetson P.B., Harris W.E., 1988, \aj\ 96, 909
\bibitem{ref.47} Stetson P.B., VandenBerg D.A., Bolte M.,
Hesser J.E., Smith G.H., 1989, \aj\ 97, 1360
\bibitem{ref.48} Stetson P.B., VandenBerg D.A., Bolte M., 1996, PASP 108, 560
\bibitem{ref.49} Storm J., Carney B.W., Latham D.W., 1994, \aap\ 290, 443
\bibitem{ref.50} Straniero O., 1988, A\&AS 76, 157
\bibitem{ref.51} Straniero O., Chieffi A., 1991, \apjs\ 76, 525
\bibitem{ref.52} VandenBerg D.A., Bolte M., Stetson P.B., 1990, \aj\
100, 445
\bibitem{ref.53} Walker A.R., 1992a, \apj\ 390, L81
\bibitem{ref.54} Walker A.R., 1992b, \pasp\ 104, 1063
\bibitem{ref.55} Walker A.R., 1994, \aj\ 108, 555
\bibitem{ref.56} Zinn R., 1985, \apj\ 293, 424
\bibitem{ref.57} Zinn R., West M.J., 1984, \apjs\ 55, 45
\end{thebibliography}
\end{document}